\documentclass[10pt]{article}

\usepackage{ajkslim}

\usepackage[authoryear,round]{natbib}
\setcitestyle{aysep={,}}
\usepackage{caption}
\microtypesetup{expansion=false}
\usepackage{newtxtext,newtxmath}
\usepackage{xspace}
\usepackage{tabularx}
\usetikzlibrary{patterns,arrows.meta}
\newcommand{\degC}{\textdegree C\xspace}
\newcommand{\degF}{\textdegree F\xspace}
\newcommand{\new}[1]{#1}
\graphicspath{{./}}
\makeatletter
\renewcommand{\section}{\@startsection{section}{1}{\z@}{-2.2ex plus -1ex minus -.2ex}{1ex plus .2ex}{\normalfont\large\bfseries\MakeUppercase}}
\renewcommand{\subsection}{\@startsection{subsection}{2}{\z@}{-1.8ex plus -1ex minus -.2ex}{0.8ex plus .2ex}{\normalfont\normalsize\bfseries}}
\makeatother
\fancypagestyle{plain}{\fancyhf{}\rfoot{\thepage}}
\crefname{figure}{Figure}{Figures}
\Crefname{figure}{Figure}{Figures}
\crefname{table}{Table}{Tables}
\crefname{equation}{Equation}{Equations}

\newcommand{\emeas}{e_{\mathrm{meas}}}
\newcommand{\eratio}{e_{\mathrm{limit}}}
\newcommand{\eexact}{e_{\mathrm{exact}}}
\newcommand{\Stemp}{S_{\mathrm{temp}}}
\newcommand{\Sshift}{S_{\mathrm{shift}}}

\title{\textbf{\new{Adjacent Heat Transfer Effects Can Bias Savings Estimates When Controlling Only Part of a Multi-Zone Building}}}
\author{\textbf{Arash J. Khabbazi}\\ \textit{Student Member ASHRAE} \and \textbf{Kevin J. Kircher}\\ \textit{Full Member ASHRAE}}
\date{}

\begin{document}
\maketitle
\thispagestyle{plain}

\vspace{-2ex}\noindent\small Arash J. Khabbazi is a PhD candidate and Kevin J. Kircher is an assistant professor of Mechanical Engineering at Purdue University.\normalsize
\section*{Abstract}
\itshape
Field demonstrations of advanced \new{heating, ventilation, and air-conditioning} control often operate a few thermal zones within a larger building and report savings for those zones alone. A recent theorem showed that this practice overestimates whole-building savings, because heat drawn from the neighboring zones is paid for by their equipment and is not counted. The error grows with the thermal conductance between the controlled and uncontrolled zones, with how much colder the controlled zones run over the test, and with the price of the heat their neighbors supply. For one controlled zone with one neighbor, a single thermal price, and slowly changing prices, the relative error reduces to the ratio of two conductances, a screening number that can be evaluated from the two conductances before a demonstration begins. This paper tests how far that reduction can be taken. A two-story house with one thermal zone per floor is simulated over a 90-day heating season in West Lafayette, Indiana, with receding-horizon model predictive control on the first floor and setpoint control on the second, under four combinations of electricity tariff (time-of-use or flat) and heating plant (heat pump or electric resistance). The exact correction from the theorem recovers the true savings in every case, and the full theorem, evaluated from the controlled zone's temperatures, the prices, and the building parameters, reproduces the measured relative error in every case. The conductance ratio bounds that error from above in all four cases, tightly under a flat tariff, within 3\%, and loosely under a time-of-use tariff, where most of the controller's savings come from shifting load through the thermal mass rather than from lowering the temperature.
\normalfont

\section{Introduction}
\new{Recent reviews} of field demonstrations of model predictive control (MPC) and reinforcement learning (RL) for building \new{heating, ventilation, and air-conditioning (HVAC) \citep{khabbazi2024,khabbazi2025}} found that advanced control can deliver real savings in occupied buildings, and also found a systematic weakness in how those savings are measured. Many studies control a few thermal zones inside a larger building and report energy or cost savings for the controlled zones only\new{, and such studies report substantially higher savings than whole-building tests. The reviews attributed the gap largely to heat transfer between the controlled zones and the adjacent uncontrolled zones. In heating, for example, an advanced controller may lower the temperature of a controlled zone, and heat then flows into it from the adjacent zones, which puts an additional heating burden on the adjacent zones. This extra heat is not counted when savings are reported for the controlled zones alone. In measurement and verification (M\&V), energy effects of a measure that cannot be measured within the measurement boundary are called interactive effects, and their magnitude needs to be estimated or evaluated to determine savings \citep{evo2022}. When the boundary encloses only the controlled zones, as in a retrofit-isolation approach \citep{ashrae2023}, this extra heating energy is such an effect. It can be captured by drawing the boundary around the whole building, or estimated with a model of the coupling between zones.} Advanced HVAC control has been adopted slowly in practice (\new{\citealp{blum2022};} \citealp{henze2024}), and trustworthy field demonstrations, with sound \new{M\&V}, are one way to change that \citep{pergantis2024}. Early demonstrations recognized the cross-zone issue and worked around it \citep{ruud1990,braun2002,braun2003,henze2005,liu2006}; more recent work has largely not.

The authors' earlier work \citep{khabbazi2026acc} traced this error to its source. \new{That work models a building as a thermal network with any number of nodes, any subset of which is under advanced control. A node can represent the air in a room or the thermal mass of walls, floors, and furnishings, and the connections between nodes can represent conduction, convection, or radiation.} For buildings whose thermal dynamics are linear and whose cost is linear in thermal load\new{, with thermal prices that may change arbitrarily in time}, that work characterized the overestimation exactly, as a conductance-weighted and price-weighted integral of the temperature reduction in the controlled zones \new{(\cref{eq:Egen} below)}, and showed that the true whole-building savings can be recovered from zone temperatures and conductance estimates, with no energy baseline. It also gave a simpler special case. In the limit of slowly changing prices, the relative overestimation for a single controlled zone with one adjacent zone and one thermal price reduces to the ratio of the conductance to the adjacent zone over the conductance to the outdoors, so a well-insulated exterior and a poorly insulated partition make the reported savings mostly fictitious, and a purely interior zone makes them entirely so. \new{The general result is hard to apply directly, because a real building has many conductances that cannot all be estimated. The special case is easy to apply, but it rests on assumptions that a real demonstration may violate.}

Two questions follow. Does the ratio remain a usable prediction over a whole heating season, for a zone with the thermal mass of a floor of a house, under an actual utility tariff? And outside its limit, how far is it from the measured error, and what determines the gap? These are practical questions. The ratio needs no \new{energy or temperature} measurements and can be evaluated before a demonstration begins, whereas the exact correction needs temperature logs, prices, and conductance estimates.
This paper answers both questions with a season-long simulation. A two-story house, with one zone per floor, is heated through December, January, and February in West Lafayette, Indiana. A receding-horizon MPC operates the first floor while the second floor holds its setpoint, in four cases that cross two tariffs\new{, time-of-use (TOU) and flat,} with two heating plants. The contributions are as follows.
\begin{itemize}[nosep]
\item The exact correction recovers the true savings in every case, and the full theorem, evaluated from the controlled floor's temperatures, the prices, and the building parameters, predicts the measured error in every case, under both tariffs and both plants.
\item The conductance ratio bounds the relative error from above in all four cases, tightly when the savings come from running the zone colder and loosely when they come from shifting load through the thermal mass. The gap between the two is set by one quantity, the ratio of load-shifting savings to lower-temperature savings, which is small when the price is flat and large when the price steps and the zone has thermal mass to shift load through.
\end{itemize}
\section{\new{Background}}
\begin{figure}[!tbp]
\centering
\resizebox{0.78\textwidth}{!}{%
\begin{tikzpicture}[x=1cm,y=1cm,line width=0.9pt,font=\large,
  res/.style={decorate,decoration={zigzag,segment length=2.2pt,amplitude=1.6pt,pre length=3pt,post length=3pt}},
  dot/.style={circle,fill,inner sep=1.3pt}]
\begin{scope}[shift={(0,0)}]
  \draw (-1.6,0) -- (6.6,0);
  \foreach \x in {-1.6,6.6} {\draw (\x,0) -- (\x,-0.25); \draw (\x-0.3,-0.25) -- (\x+0.3,-0.25); \draw (\x-0.18,-0.4) -- (\x+0.18,-0.4); \draw (\x-0.06,-0.55) -- (\x+0.06,-0.55);}
  \node[dot,label=above left:{$T_1$}] (T1) at (0,2.4) {};
  \draw (T1) -- (0,1.55); \draw (-0.45,1.55) -- (0.45,1.55); \draw (-0.45,1.3) -- (0.45,1.3); \draw (0,1.3) -- (0,0);
  \node[right] at (0.5,1.42) {$C_1$};
  \draw (T1) -- (-1.6,2.4) -- (-1.6,1.75);
  \draw (-1.6,1.25) circle (0.5); \draw[-{Latex[length=5pt]}] (-1.6,0.95) -- (-1.6,1.55);
  \draw (-1.6,0.75) -- (-1.6,0);
  \node[left] at (-2.15,1.25) {$q_1{+}w_1$};
  \node[dot,label=above right:{$T_2$}] (T2) at (5,2.4) {};
  \draw (T2) -- (5,1.55); \draw (4.55,1.55) -- (5.45,1.55); \draw (4.55,1.3) -- (5.45,1.3); \draw (5,1.3) -- (5,0);
  \node[left] at (4.5,1.42) {$C_2$};
  \draw (T2) -- (6.6,2.4) -- (6.6,1.75);
  \draw (6.6,1.25) circle (0.5); \draw[-{Latex[length=5pt]}] (6.6,0.95) -- (6.6,1.55);
  \draw (6.6,0.75) -- (6.6,0);
  \node[right] at (7.15,1.25) {$q_2{+}w_2$};
  \draw (T1) -- (1.4,2.4); \draw[res] (1.4,2.4) -- (3.6,2.4); \draw (3.6,2.4) -- (T2);
  \node[above] at (2.5,2.55) {$R_{12}$};
  \draw (T1) -- (0,2.9); \draw[res] (0,2.9) -- (0,4.1); \draw (0,4.1) -- (0,4.5);
  \node[left] at (-0.15,3.5) {$R_{10}$};
  \draw (T2) -- (5,2.9); \draw[res] (5,2.9) -- (5,4.1); \draw (5,4.1) -- (5,4.5);
  \node[right] at (5.15,3.5) {$R_{20}$};
  \draw (0,4.5) -- (5,4.5);
  \node[dot,label=above:{$T_0$ (outdoors)}] at (2.5,4.5) {};
\end{scope}
\begin{scope}[shift={(10.3,0)}]
  \def\W{4.2}\def\H{1.5}
  \node[align=center] at (-1.35,2.2) {Zone 0\\(outdoors)};
  \fill[pattern=north east lines,pattern color=gray] (0,-0.3) rectangle (\W,0);
  \draw[gray] (0,-0.3) rectangle (\W,0);
  \node[below,gray] at (\W/2,-0.3) {adiabatic slab};
  \draw[blue,line width=1.6pt] (0,0) -- (0,2*\H) -- (\W,2*\H) -- (\W,0);
  \draw[blue,line width=1.6pt] (0,2*\H) -- (\W/2,2*\H+0.9) -- (\W,2*\H);
  \draw[red,line width=1.6pt] (0,\H) -- (\W,\H);
  \node[align=center] at (\W/2,1.5*\H) {Floor 2, zone 2\\(default control)};
  \node[align=center] at (\W/2,0.5*\H) {Floor 1, zone 1\\(advanced control)};
  \node[red,right] at (\W+0.1,\H) {$U_{\mathrm{int}}$};
  \node[blue,right] at (\W+0.1,0.5*\H) {$U_{\mathrm{ext}}$};
  \node[blue,above right] at (\W/2+0.9,2*\H+0.75) {$U_{\mathrm{ext}}$};
  \draw[{Bar}-{Bar}] (-0.35,0) -- (-0.35,\H); \node[left] at (-0.4,\H/2) {$h$};
\end{scope}
\end{tikzpicture}}
\caption{The left panel shows the thermal circuit for the two-zone building, with $R_{i0} = 1/\alpha_{i0}$ and $R_{12} = 1/\alpha_{12}$. The right panel shows the two-story house in section. The advanced controller operates floor~1 only; floor~2 stays under default control. The exterior envelope (blue, $U_{\mathrm{ext}}$) is better insulated than the interfloor assembly (red, $U_{\mathrm{int}}$); the slab is adiabatic.}
\label{fig:model}
\end{figure}
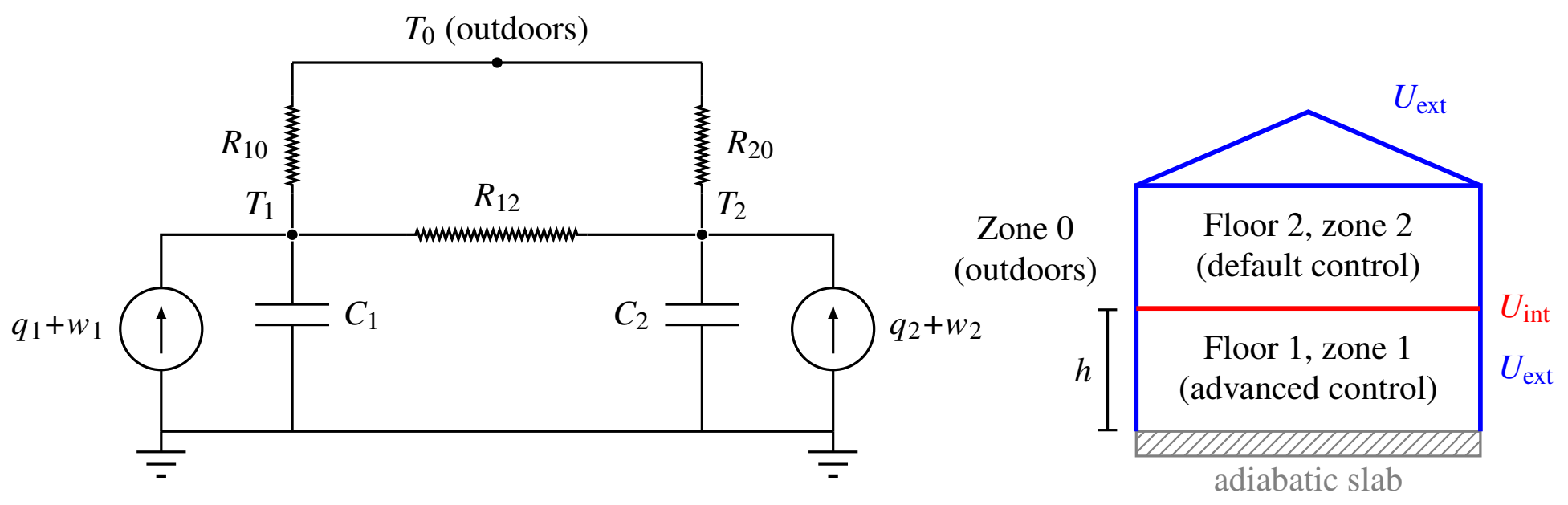

This section summarizes the parts of the earlier work \citep{khabbazi2026acc} that the simulation uses; \new{the proof is not repeated. Consider a building with $n$ thermal zones, of which zones $1, \dots, m$ are under advanced control and zones $m+1, \dots, n$ are not, and let zone~0 denote the outdoors.} Zone $i$ has thermal capacitance $C_i$ (kWh/\degC) \new{and exchanges heat with zone~$j$ through the conductance $\alpha_{ij}$} (kW/\degC). Heating equipment delivers thermal power $q_i(t)$ (kW) and exogenous sources deliver $w_i(t)$. Two scenarios are compared, a \emph{baseline}, in which a benchmark controller holds \new{every zone} at \new{its setpoint}, and an \emph{experiment}, in which an advanced controller takes over \new{zones $1, \dots, m$} while \new{the other zones remain} under the benchmark controller. Quantities in the experiment carry a tilde\new{, so $T_i(t)$ and $\tilde T_i(t)$ are the temperatures of zone~$i$ in the baseline and the experiment}. Delivering $q_i$ costs $a_i(t)\,q_i(t)$ per unit time, where $a_i$ is a thermal price; for electric heating, $a = \pi/\eta$, the electricity price divided by the coefficient of performance (COP). \new{For a test of duration $\tau$, the earlier work shows that reporting the savings of the controlled zones alone overestimates the whole-building savings by
\begin{equation}
E = \sum_{i=1}^{m} \sum_{j=m+1}^{n} \alpha_{ij} \int_0^\tau a_j(t)\,\bigl(T_i(t) - \tilde T_i(t)\bigr)\,dt .
\label{eq:Egen}
\end{equation}
In heating, each term is the cost of the extra heat that flows from an uncontrolled zone~$j$ into a controlled zone~$i$ when zone~$i$ runs colder than in the baseline, priced at $a_j$ because the equipment in zone~$j$ supplies it. The simulation uses the simplest case, $n = 2$ and $m = 1$, with conductances $\alpha_{10}$ and $\alpha_{20}$ to the outdoor temperature $T_0$ and $\alpha_{12}$ between the two zones.} The prices are assumed spatially uniform, $a_1 = a_2 = a$, which holds when both zones share one tariff and one type of plant. \Cref{fig:model} shows the circuit and the building it represents.

Let $x(t) = T_1(t) - \tilde T_1(t)$ be the temperature reduction in zone~1 relative to the baseline, and define
\begin{equation}
A = \int_0^\tau a(t)\,x(t)\,dt, \qquad B = -\int_0^\tau \frac{da}{dt}\,x(t)\,dt .
\label{eq:AB}
\end{equation}
For this two-zone case, \new{\cref{eq:Egen} reduces to}
\begin{equation}
E = \alpha_{12}\,A,
\label{eq:E}
\end{equation}
and \new{the earlier work shows} that the true whole-building savings are
\begin{equation}
S = \alpha_{10}\,A + C_1\,B ,
\label{eq:S}
\end{equation}
provided the two scenarios begin and end at the same zone-1 temperature. The perceived savings are $S_1 = E + S$. Two corrections follow, labeled I and II. A practitioner who has measured $S_1$ can subtract the error, $\hat S_{\mathrm{I}} = S_1 - \hat\alpha_{12} A$, which needs one conductance estimate. Alternatively, $\hat S_{\mathrm{II}} = \hat\alpha_{10} A + \hat C_1 B$ needs no energy measurement in zone~1 at all, only its temperature, the prices, and two parameter estimates. A hat denotes an estimate, so $\hat\alpha_{12}$, $\hat\alpha_{10}$, and $\hat C_1$ are the parameter values a practitioner has, and $\hat S_{\mathrm{I}}$ and $\hat S_{\mathrm{II}}$ are the corrected savings they give; with the true parameters both equal $S$.

Define the relative overestimation that the full theorem predicts from the controlled zone's temperatures, the prices, and the building parameters as
\begin{equation}
\eexact = \frac{\alpha_{12}\,A}{\alpha_{10}\,A + C_1\,B} .
\label{eq:eexact}
\end{equation}
By \cref{eq:E,eq:S}, the theorem asserts that the measured relative error $E/S$ equals $\eexact$ for any price trajectory, so the error can be predicted without measuring any energy. In the limit of slowly changing prices, $da/dt \to 0$, $B$ vanishes and \cref{eq:eexact} reduces to
\begin{equation}
\eratio = \frac{\alpha_{12}}{\alpha_{10}} ,
\label{eq:epred}
\end{equation}
independent of temperatures, prices, and the controller. This is the screening result.

One exact consequence of \cref{eq:E,eq:S} is used throughout. The two terms of \cref{eq:S} have distinct physical meanings. The first, $\Stemp = \alpha_{10} A$, is what running zone~1 colder saves through reduced envelope loss. The second, $\Sshift = C_1 B$, is what the controller earns by storing heat in the zone's thermal mass when the thermal price is low and coasting when it is high; it is nonzero only if the price changes in time and the zone has mass to store heat in. Writing $r = \Sshift/\Stemp$ and dividing \cref{eq:E} by \cref{eq:S}, the measured relative error, the ratio of the two savings from the heating loads, obeys
\begin{equation}
\emeas = \frac{E}{S} = \frac{\eratio}{1 + r}, \qquad r = \frac{C_1 B}{\alpha_{10} A} = \frac{C_1}{\alpha_{10}} \cdot \frac{B}{A} .
\label{eq:emeas}
\end{equation}

\section{Simulation Study}
\subsection{Building}
The building is a representative two-story detached house, not a specific one\new{, with one thermal zone per floor and the dimensions and envelope properties listed in \cref{tab:params}. The first floor's slab is treated as adiabatic, the second floor carries the roof, and the interfloor assembly has the full footprint area. The heat transfer coefficients follow \citet{khabbazi2026acc}, with the exterior envelope better insulated than the interfloor assembly, and the resulting slow-price limit is} $\eratio = \alpha_{12}/\alpha_{10} = 1.905$.

Each zone's capacitance is the heat capacity of its air multiplied by 30 to represent furnishings, interior surfaces, and structure in close thermal contact with the air\new{, which gives the first floor a time constant of 6.4~h}. \new{The resulting capacitance of the whole house, 6.3~kWh/\degC\ (12{,}000~Btu/\degF), is close to the roughly 6.4~kWh/\degC\ (12{,}100~Btu/\degF) implied by the model identified from field data for a 208~m$^2$ (2239~ft$^2$) two-story test house in West Lafayette in which MPC has been demonstrated \citep{pergantis2024humidity}.} Nothing is fitted to measurements from any building.

\begin{table}[!tbp]
\centering
\caption{Building, Plant, and Study Parameters}
\label{tab:params}
\footnotesize
\setlength{\tabcolsep}{4pt}
\begin{tabularx}{\textwidth}{@{}>{\raggedright\arraybackslash}p{0.2\textwidth}>{\raggedright\arraybackslash}X@{\hspace{1em}}>{\raggedright\arraybackslash}p{0.2\textwidth}>{\raggedright\arraybackslash}X@{}}
\toprule
Quantity & Value & Quantity & Value \\
\midrule
Footprint, ceiling height & 9 m $\times$ 12 m, 2.7 m \new{(29.5 ft $\times$ 39.4 ft, 8.9 ft)} & Internal gains, solar aperture & 0.3 kW \new{(1020 Btu/h)}; 1.5 m$^2$ \new{(16.1 ft$^2$)} per floor \\
Exterior area, floor 1 / 2 & 113 (walls) / 221 (walls, roof) m$^2$ \new{(1216 / 2379 ft$^2$)} & Heat pump COP & 1.8 at $-15$\degC\ \new{(5\degF)}, 3.3 at 8.3\degC\ \new{(47\degF)}, linear \\
Interfloor area & 108 m$^2$ \new{(1163 ft$^2$)} & Heating capacity, floor 1 / 2 & 8.25 / 16.4 kW \new{(28{,}200 / 56{,}000 Btu/h)} \\
$U$, exterior / interfloor & 1.5 / 3.0 W/(m$^2\cdot$\degC) \new{[0.26 / 0.53 Btu/(h$\cdot$ft$^2\cdot$\degF)]} & TOU price, off / peak / mid (\$/kWh) & 0.086 / 0.211 / 0.143 \\
$\alpha_{10}$, $\alpha_{20}$, $\alpha_{12}$ & 0.170, 0.332, 0.324 kW/\degC\ \new{[322, 629, 614 Btu/(h$\cdot$\degF)]} & TOU hours, off / peak & 0--4 h / 6--8 and 17--21 h \\
$C_1 = C_2$ & 3.16 kWh/\degC\ \new{(5990 Btu/\degF)} (air $\times$ 30) & Flat price & \$0.13/kWh \\
Setpoint, comfort band & 21\degC\ \new{(70\degF)}; $\pm$1\degC\ \new{($\pm$1.8\degF)} 6--9, 18--22 h; $\pm$2\degC\ \new{($\pm$3.6\degF)} otherwise & Season; time step; MPC horizon & Dec 1 to Feb 28; 1 h; 24 h \\
\bottomrule
\end{tabularx}
\end{table}

\subsection{Weather, Gains, and Comfort}
\begin{figure}[!tbp]
\centering
\includegraphics[width=0.55\textwidth]{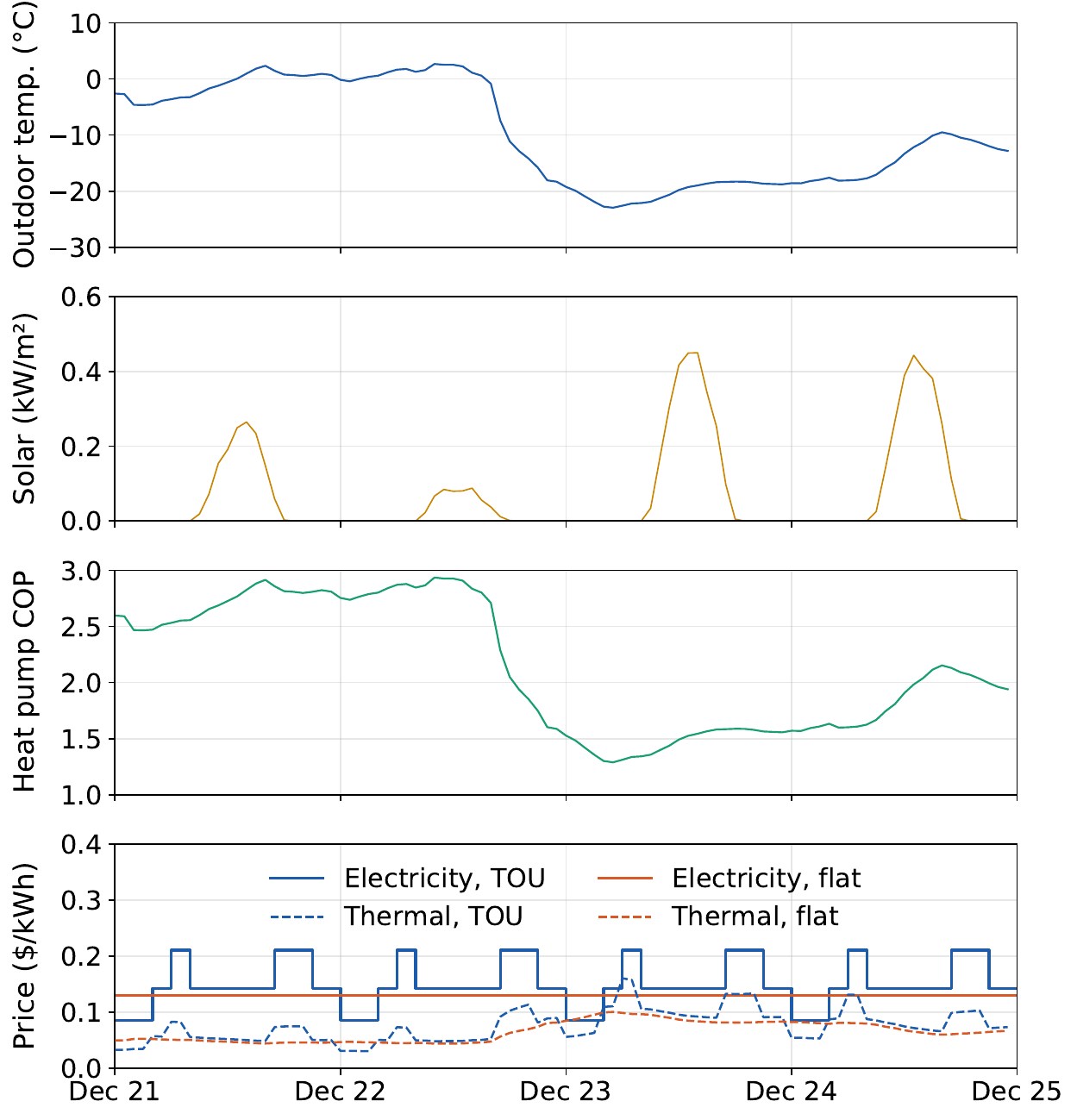}
\caption{Simulation inputs over the four coldest days, December 21 to 24. Outdoor temperature, global horizontal irradiance, heat pump COP, and prices. In the price panel the electricity price is solid and the heat pump's thermal price is dashed, for the TOU and flat tariffs. With resistance heat the thermal price equals the electricity price.}
\label{fig:inputs}
\end{figure}

Outdoor temperature and global horizontal irradiance are hourly ERA5 reanalysis values \citep{hersbach2020} for West Lafayette from December 1, 2022 to February 28, 2023. The outdoor temperature averages 1.8\degC\ \new{(35.2\degF)} and falls to $-22.9$\degC\ \new{($-9.2$\degF)} during a cold snap on December 22 to 24. Each floor receives constant internal gains plus solar gains through an effective aperture, for a peak of 1.35~kW \new{(4610~Btu/h)}, \new{chosen} so that no hour of the season requires cooling. The comfort band around the setpoint is \new{narrower} while occupants are awake at home. \Cref{fig:inputs} shows the inputs over the four coldest days, December 21 to 24, the window used for the results figure.

\subsection{Heating Plant and Tariffs}
Each floor has its own heating plant, sized to 115\% of that floor's peak baseline load. The heat pump's COP varies linearly with outdoor temperature between the two rating points in \cref{tab:params} and ranges from 1.29 to 3.83 over the season, and electric resistance heat has a COP of one. No backup resistance heat is added to the heat pump, because \new{the theorem assumes a cost that is linear in load}.

Two tariffs from a large Indiana utility are used, as documented by \citet{dierking2026}, a \new{TOU} rate with off-peak, peak, and mid-peak prices and a flat rate, both listed in \cref{tab:params}. Crossing the two plants with the two tariffs gives four cases, labeled HP-TOU, HP-Flat, Res-TOU, and Res-Flat. The cases differ in how the thermal price $a = \pi/\eta$ moves. It steps daily and drifts with outdoor temperature in HP-TOU, drifts only in HP-Flat, steps only in Res-TOU, and is constant in Res-Flat. The last case is the regime in which the slow-price limit holds by construction.

\subsection{Controller}
The advanced controller on the first floor is a receding-horizon MPC \citep{serale2018,drgona2020}. Every hour it solves a convex program over the next 24~h, applies the first hour's heating decision, and re-solves from the resulting state. The plant model in the controller is the same first-order zone model as the simulation, discretized exactly under a zero-order hold, with perfect foresight of weather, prices, and gains; the second floor's temperature enters as a known input. The objective is the cumulative electricity cost, $\sum_k \pi(k)\,\tilde q_1(k)/\eta(k)$, subject to the comfort band, a nonnegative heating rate below the plant's capacity, and a terminal condition that the zone ends each horizon at or above its setpoint. Every solve that can see the end of the season pins the first floor to its setpoint at the final hour, so the two scenarios begin and end at the same temperature, which \cref{eq:S} requires. The convex programs are solved with CVXPY \citep{diamond2016}. The second floor holds its setpoint exactly in both scenarios; its heating rate in the experiment rises by $\alpha_{12}$ times the first floor's temperature reduction.

Under a flat tariff the cost-optimal policy is simple\new{. It holds} the bottom of the comfort band and \new{recovers} just before an occupied period. The same controller is used in all four cases for consistency, but in the flat-rate cases it reduces to a scheduled setback, with a small anticipatory preheat before cold spells in HP-Flat because the COP is about to drop. Those cases are included not to show MPC being useful but because a setback thermostat on a flat rate is the most common configuration in residential demonstrations, and because they sit in the regime where \cref{eq:epred} is expected to hold.

\subsection{Savings Accounting}
Each case is simulated twice, once with both floors on setpoint and once with the MPC on the first floor. Costs are summed per floor and per scenario. Perceived savings are the first floor's cost reduction; true savings are the whole house's. The integrals in \cref{eq:AB} are evaluated as hourly sums. The integral $A$ of \cref{eq:AB} uses $x$ averaged exactly over each hour from the zero-order-hold solution. For the integral $B$ of \cref{eq:AB}, the derivative of a stepped tariff is a delta function at every step, so the derivative is moved onto the temperature instead. Integrating by parts with matched end states,
\begin{equation}
B = -\int_0^\tau \frac{da}{dt}\,x\,dt = \int_0^\tau a\,\frac{dx}{dt}\,dt = \sum_k a_k\,(x_{k+1} - x_k) = -\sum_k (a_{k+1} - a_k)\,x_{k+1},
\label{eq:Bdisc}
\end{equation}
where the second equality is the identity in the appendix of \citet{khabbazi2026acc}, the third uses a price constant within each hour, and the last is summation by parts. The final form shows that only the hours at which the price jumps contribute, each by the jump times $x$ at that instant, so the tariff steps are captured exactly and the price is never differentiated. The continuous COP variation enters through the same sum. \Cref{eq:E,eq:S,eq:emeas} are then checked against the simulated costs in every case.

\section{Results}
\subsection{Season Accounting}
\Cref{tab:acc} reports the season's costs and savings for each case. In HP-TOU, the first floor's heating cost falls from \$350 to \$247, a perceived saving of \$103. The second floor's cost rises from \$731 to \$771 because it makes up the heat that now flows downward through the interfloor assembly. The house as a whole saves \$62. Looking at the first floor alone would overstate the savings by 65\%. The same pattern holds in every case; the dollar totals of the resistance cases are larger because each kilowatt-hour of heat costs roughly three times more, so comparisons across cases should be made on percentages and ratios rather than on dollars.

\begin{table}[!tbp]
\centering
\caption{Energy Costs and Savings over the Season, by Case (\$)}
\label{tab:acc}
\footnotesize
\begin{tabular}{lrrrrrrrr}
\toprule
& \multicolumn{2}{c}{HP-TOU} & \multicolumn{2}{c}{HP-Flat} & \multicolumn{2}{c}{Res-TOU} & \multicolumn{2}{c}{Res-Flat} \\
\cmidrule(lr){2-3}\cmidrule(lr){4-5}\cmidrule(lr){6-7}\cmidrule(lr){8-9}
& Floor 1 & Floor 2 & Floor 1 & Floor 2 & Floor 1 & Floor 2 & Floor 1 & Floor 2 \\
\midrule
Baseline cost & 350.25 & 730.65 & 304.28 & 634.78 & 920.66 & 1930.45 & 798.25 & 1673.91 \\
Experiment cost & 247.23 & 771.29 & 221.32 & 688.67 & 639.85 & 2048.08 & 568.34 & 1824.67 \\
Perceived zone-level savings & 103.02 & $-$40.64 & 82.95 & $-$53.89 & 280.81 & $-$117.63 & 229.91 & $-$150.76 \\
True savings, whole building & \multicolumn{2}{c}{62.38} & \multicolumn{2}{c}{29.06} & \multicolumn{2}{c}{163.18} & \multicolumn{2}{c}{79.15} \\
\bottomrule
\end{tabular}
\end{table}

\subsection{The Coldest Days}
\begin{figure}[!tbp]
\centering
\includegraphics[width=0.55\textwidth]{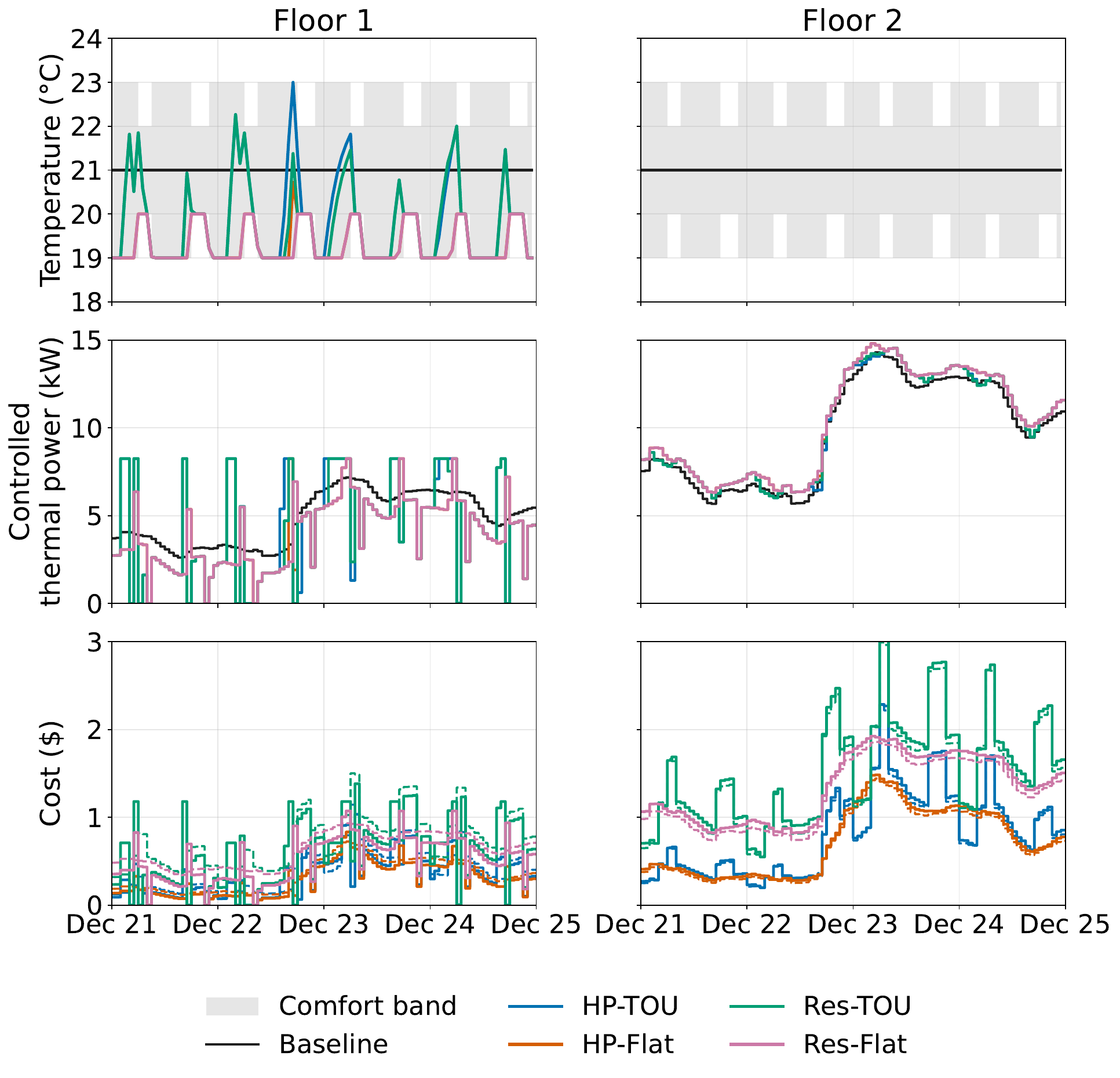}
\caption{The four coldest days, December 21 to 24, all cases. Zone temperatures with the comfort band (top), controlled thermal power (middle), and hourly cost (bottom) for the first floor under the advanced controller (left) and the second floor under setpoint control (right). In the cost row each case's baseline is dashed and its experiment solid. The second floor's temperature is the same in every scenario and case, so only one line appears in its top panel. Totals for the window are in \cref{tab:accweek}.}
\label{fig:week}
\end{figure}

\Cref{fig:week} shows the four coldest days, December 21 to 24, for all four cases on one time axis, with the first floor on the left and the second on the right. The cases pair off by tariff. Under the TOU rate the controller holds the first floor at the bottom of the band during cheap hours, drives it to the top of the band before each morning and evening price step, and coasts through the expensive period; the heat-pump and resistance traces lie on top of each other and separate only before the coldest morning, when the heat pump preheats slightly more ahead of the drop in its COP. Under the flat rate the controller sits at the bottom of the band and rises to the occupied band edge on schedule, with one small anticipatory preheat on the heat pump before the outdoor temperature drops on December 22. Over the whole season the two plants' trajectories differ by 0.03\degC\ \new{(0.05\degF)} on average under either tariff.

In the middle row, the first floor alternates between full capacity and near zero under the TOU rate and runs below its baseline in nearly every hour under the flat rate, while the second floor runs above its baseline in every hour of every case, by an amount proportional to how far the first floor sits below setpoint. The bottom row converts both to cost\new{, and} the second floor's experiment runs above its baseline in every case. \Cref{tab:accweek} gives the totals of the same four days in the same form as \cref{tab:acc}. The second floor's penalty is the overestimation that a first-floor-only report would miss; over these four days it is 39\% of the perceived savings in HP-TOU and 66\% in Res-Flat.

\begin{table}[!tbp]
\centering
\caption{Energy Costs and Savings over the Four Coldest Days, December 21 to 24, by Case (\$)}
\label{tab:accweek}
\footnotesize
\begin{tabular}{lrrrrrrrr}
\toprule
& \multicolumn{2}{c}{HP-TOU} & \multicolumn{2}{c}{HP-Flat} & \multicolumn{2}{c}{Res-TOU} & \multicolumn{2}{c}{Res-Flat} \\
\cmidrule(lr){2-3}\cmidrule(lr){4-5}\cmidrule(lr){6-7}\cmidrule(lr){8-9}
& Floor 1 & Floor 2 & Floor 1 & Floor 2 & Floor 1 & Floor 2 & Floor 1 & Floor 2 \\
\midrule
Baseline cost & 37.97 & 76.95 & 32.92 & 66.73 & 69.72 & 141.68 & 60.22 & 122.41 \\
Experiment cost & 31.40 & 79.49 & 27.60 & 70.15 & 57.09 & 147.12 & 49.76 & 129.26 \\
Perceived zone-level savings & 6.57 & $-$2.54 & 5.31 & $-$3.42 & 12.63 & $-$5.44 & 10.45 & $-$6.85 \\
True savings, whole building & \multicolumn{2}{c}{4.03} & \multicolumn{2}{c}{1.90} & \multicolumn{2}{c}{7.19} & \multicolumn{2}{c}{3.60} \\
\bottomrule
\end{tabular}
\end{table}

\subsection{Where the Savings Come From}
\Cref{tab:decomp} splits the season's results using \cref{eq:E,eq:S}. The first row is the overestimation error, which the simulation gives as perceived minus true savings and the theorem gives as $\alpha_{12} A$; the two agree exactly. The next two rows are the two terms of the true savings. In the flat-rate cases nearly all of the savings come from running the first floor colder. The load-shifting term is \$0.77 for the heat pump, from the small COP-driven preheats, and zero for resistance heat. In the TOU cases the picture reverses. \new{About two} thirds of the true savings come from load shifting, \$41 of \$62 for the heat pump and \$101 of \$163 for resistance, and $r$ is 1.9 and 1.6.

\begin{table}[!tbp]
\centering
\caption{Overestimation and the Two Components of True Savings, by Case}
\label{tab:decomp}
\footnotesize
\begin{tabular}{p{0.5\textwidth}rrrr}
\toprule
& HP-TOU & HP-Flat & Res-TOU & Res-Flat \\
\midrule
Overestimation error $E = \alpha_{12} A$ (\$) & 40.64 & 53.89 & 117.63 & 150.76 \\
Lower-temperature term $\Stemp = \alpha_{10} A$ (\$) & 21.33 & 28.29 & 61.76 & 79.15 \\
Load-shifting term $\Sshift = C_1 B$ (\$) & 41.05 & 0.77 & 101.42 & 0.00 \\
$r = \Sshift / \Stemp$ & 1.924 & 0.027 & 1.642 & 0.000 \\
Overestimation, \% of true savings & 65.1 & 185.5 & 72.1 & 190.5 \\
$\emeas = E/S$, from the heating loads & 0.651 & 1.855 & 0.721 & 1.905 \\
$\eexact$, full theorem from temperatures and prices, \cref{eq:eexact} & 0.651 & 1.855 & 0.721 & 1.905 \\
$\eratio = \alpha_{12}/\alpha_{10}$, slow-price limit of the theorem & 1.905 & 1.905 & 1.905 & 1.905 \\
\bottomrule
\end{tabular}
\end{table}

The last three rows are the paper's central result. The full theorem, evaluated from the first floor's temperatures and the prices with no energy data, reproduces the measured ratio in every case; the two rows agree very closely because the plant obeys the model and the end states match, and the agreement is the check that the accounting and \cref{eq:Bdisc} are right. The slow-price limit is 1.905 in every case. The measured ratio is 1.905 in Res-Flat, where the thermal price is constant and the end states are matched, so $B = 0$ and \cref{eq:epred} holds exactly. It is 1.855 in HP-Flat, 3\% below the ratio, the whole of the gap coming from the COP's dependence on outdoor temperature. It is 0.72 in Res-TOU and 0.65 in HP-TOU, \new{about} one third of the slow-price limit. \Cref{eq:emeas} accounts for all four values, for example $1.905/(1 + 1.924) = 0.651$ for HP-TOU.

\begin{figure}[!tbp]
\centering
\includegraphics[width=0.68\textwidth]{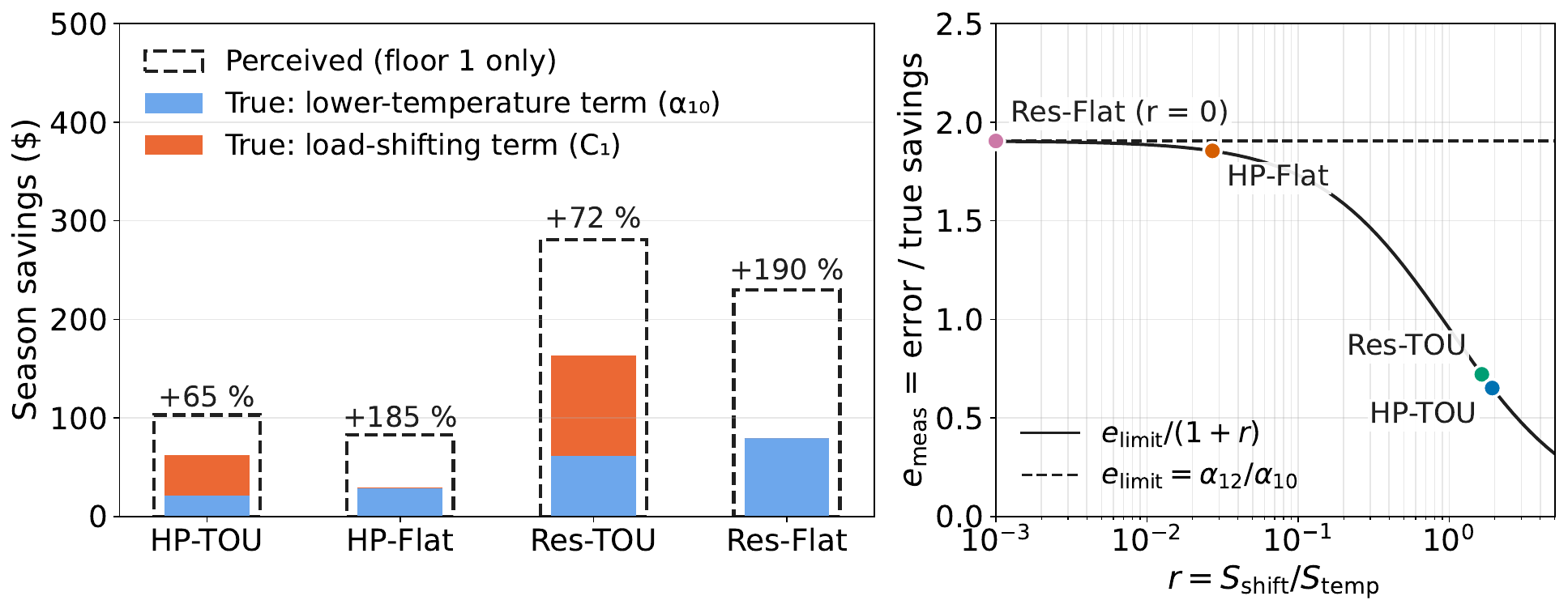}
\caption{The left panel shows the true savings by case, stacked from the lower-temperature term and the load-shifting term, with perceived savings as the dashed outline and the overestimation annotated. The right panel shows the measured relative error against $r$, with \cref{eq:emeas} as the solid curve and $\eratio$ as the dashed line. Res-Flat has $r = 0$ and is drawn at the left edge.}
\label{fig:headline}
\end{figure}

\Cref{fig:headline} shows this in one place. The left panel stacks the true savings of each case from the two terms, with the perceived savings as an outline behind them, and the right panel plots the measured ratio against $r$ with \cref{eq:emeas} drawn through the points. The flat-rate cases sit on the plateau at $\eratio$; the TOU cases sit well down the curve, and the two plants of each tariff sit near each other.

\section{Discussion}
\new{These results have three implications for M\&V in small HVAC control demonstrations in larger buildings. The exact corrections are sufficient on their own, recovering the true savings in all four cases without an energy baseline.}

\new{The conductance ratio is best used as a screening bound. \Cref{eq:emeas} turns its validity into a computable condition, $r \ll 1$, rather than a qualitative appeal to slowly changing prices. The condition holds when the savings come from running the zone colder, as on a flat rate with either plant, and fails when the controller shifts load through the thermal mass, as MPC does on a TOU rate \citep{ham2023,zanetti2025}.} In that regime the ratio overstates the relative error by the factor $1 + r$. \new{Because it needs only conductance estimates and bounded the error from above in every case, the ratio remains useful before a demonstration, where a small value supports the design and a large one flags it.}

\new{The tariff, rather than the plant, sets the size of the gap.} The two TOU cases and the two flat-rate cases pair off in every table and figure. \new{Replacing the heat pump with resistance heat changes the dollar totals by about a factor of three but the measured ratio by at most about 10\%, whereas replacing the flat rate with the TOU rate changes the measured ratio by about a factor of three. The tariff is therefore the first factor to examine when judging whether the ratio applies.}

\section{Limitations and Future Work}
The scope of this study is the same as that of the theorem it tests. Heat transfer is linear in temperature differences and the cost is linear in thermal load, which holds for input energy, energy cost, and emissions but not for objectives such as peak demand. Backup resistance heat, which most cold-climate heat pumps carry \citep{pergantis2024}, makes the cost piecewise linear in load\new{, and a COP that depends on part-load ratio makes it nonlinear}.

Within that scope, the plant and the controller share one lumped two-zone model, so the corrections recover the true savings by construction and the substance of the results is the behavior of $r$. \new{Buoyancy-driven airflow between the floors is not modeled explicitly. It can be represented approximately as a convective conductance in parallel with conduction through the floor, which raises $\alpha_{12}$ and the overestimation error without changing the form of the results. The effect is weak when the first floor is colder than the second and strongest during a TOU preheat, when the first floor is warmer. Each floor is a single thermal node, without a slower node for the deeper structure. The theorem admits such mass nodes, but this example does not include them, and adding them is a direction for future work. This lumped approximation is accurate for windows and many uniform walls but less so for thick, well-insulated walls \citep{kircher2015}, and additional mass would mainly increase load shifting under the TOU tariff, which raises $r$. The simulation also assumes perfect forecasts of weather, prices, and gains, an adiabatic slab, no infiltration, and a second floor that holds its setpoint exactly, and it uses a single heating season of weather. Infiltration would add to the conductance to the outdoors and lower the conductance ratio.}

\new{Although the study uses two zones, the theorem holds for any number of zones. In a larger building, a controlled zone typically has more interior neighbors and less exterior surface, so the total conductance to adjacent zones grows relative to the conductance to the outdoors, and the overestimation grows with it; the two-story house is a mild case.} The conductance ratio that a practitioner would check before a demonstration requires the thermal conductances, which can be challenging to estimate accurately in a real building. How uncertainty in those parameters propagates into the corrected savings, and how to limit its effect, is an important direction for future work. The results should ultimately be validated experimentally, \new{for example in} a two-zone building in which advanced control is applied to one zone only.

\section{Conclusion}
This paper \new{extended the evaluation of} the theorem of the authors' previous work \citep{khabbazi2026acc} \new{to} a season-long simulation of a two-story house with one floor under advanced control, across four combinations of tariff and heating plant. The exact correction for cross-zone heat transfer recovered the true savings in every case, and the full theorem predicted the measured error from temperatures, prices, and building parameters in every case. The conductance ratio $\alpha_{12}/\alpha_{10}$ \new{matched the measured error within 3\%} on a flat tariff\new{, as the theorem's slow-price limit predicts, and bounded it from above on a time-of-use tariff}; the gap is the number $r$, and $\emeas = \eratio/(1+r)$ exactly.

\section*{Acknowledgments}
The authors gratefully acknowledge support for AJK from the ASHRAE Graduate Student Grant-in-Aid Award.

\section*{Nomenclature}
\small
\setlength{\tabcolsep}{3pt}
\begin{tabular}{@{}ll@{\hspace{1em}}ll@{}}
$a$ & thermal price, \$/kWh of heat, $= \pi/\eta$ & $r$ & $\Sshift/\Stemp$ \\
$A$, $B$ & price-weighted integrals of $x$, \cref{eq:AB} & $S$, $S_1$ & true and perceived savings, \$ \\
$C_i$ & thermal capacitance of zone $i$, kWh/\degC & $\Stemp$, $\Sshift$ & the two terms of $S$ in \cref{eq:S}, \$ \\
$E$ & overestimation error, \$ & \new{$T_0$;} $T_i$, $\tilde T_i$ & \new{outdoor;} zone $i$ temperature, baseline/experiment, \degC \\
$e_{\mathrm{meas}}$, $e_{\mathrm{exact}}$, $e_{\mathrm{limit}}$ & $E/S$ measured, \cref{eq:eexact}, \cref{eq:epred} & $x$ & $T_1 - \tilde T_1$, temperature reduction, \degC \\
$\eta$, $\pi$ & COP; electricity price, \$/kWh & $\alpha_{ij}$ & conductance, zone $i$ to $j$ (0 is outdoors), kW/\degC \\
$q_i$, $w_i$ & controlled, exogenous heat to zone $i$, kW & $\tau$ & duration of the test, h \\
\new{$n$, $m$} & \new{number of zones, of controlled zones} & \multicolumn{2}{l}{\new{$^*$SI units; I-P equivalents in the text and \cref{tab:params}}} \\
\end{tabular}


\normalsize

\begin{thebibliography}{99}
\bibitem[ASHRAE(2023)]{ashrae2023} \new{ASHRAE. 2023. \textit{ASHRAE Guideline 14-2023, Measurement of Energy, Demand, and Water Savings}. Peachtree Corners, GA: ASHRAE.}
\bibitem[Blum et al.(2022)]{blum2022} \new{Blum, D., Z. Wang, C. Weyandt, D. Kim, M. Wetter, T. Hong, and M.A. Piette. 2022. Field demonstration and implementation analysis of model predictive control in an office HVAC system. \textit{Applied Energy} 318:119104.}
\bibitem[Braun(2003)]{braun2003} Braun, J.E. 2003. Load control using building thermal mass. \textit{Journal of Solar Energy Engineering} 125(3):292--301.
\bibitem[Braun et al.(2002)]{braun2002} Braun, J.E., T.M. Lawrence, C.J. Klaassen, and J.M. House. 2002. Demonstration of load shifting and peak load reduction with control of building thermal mass. \textit{Proceedings of the 2002 ACEEE Summer Study on Energy Efficiency in Buildings}, Monterey, CA.
\bibitem[Diamond and Boyd(2016)]{diamond2016} Diamond, S., and S. Boyd. 2016. CVXPY: A Python-embedded modeling language for convex optimization. \textit{Journal of Machine Learning Research} 17(83):1--5.
\bibitem[Dierking et al.(2026)]{dierking2026} Dierking, W.G., A.J. Khabbazi, L.D. Reyes Premer, and K.J. Kircher. 2026. Scalable supervisory HVAC control for linear objectives. \textit{Proceedings of the 65th IEEE Conference on Decision and Control}. arXiv:2607.15867.
\bibitem[Drgo\v{n}a et al.(2020)]{drgona2020} Drgo\v{n}a, J., J. Arroyo, I. Cupeiro Figueroa, D. Blum, K. Arendt, D. Kim, E.P. Oll\'e, J. Oravec, M. Wetter, D.L. Vrabie, and L. Helsen. 2020. All you need to know about model predictive control for buildings. \textit{Annual Reviews in Control} 50:190--232.
\bibitem[EVO(2022)]{evo2022} \new{EVO. 2022. \textit{International Performance Measurement and Verification Protocol (IPMVP): Core Concepts}. Washington, DC: Efficiency Valuation Organization.}
\bibitem[Ham et al.(2023)]{ham2023} \new{Ham, S.W., D. Kim, T. Barham, and K. Ramseyer. 2023. The first field application of a low-cost MPC for grid-interactive K-12 schools: Lessons-learned and savings assessment. \textit{Energy and Buildings} 296:113351.}
\bibitem[Henze et al.(2005)]{henze2005} Henze, G.P., D.E. Kalz, S. Liu, and C. Felsmann. 2005. Experimental analysis of model-based predictive optimal control for active and passive building thermal storage inventory. \textit{HVAC\&R Research} 11(2):189--213.
\bibitem[Henze et al.(2024)]{henze2024} Henze, G.P., K.J. Kircher, and J.E. Braun. 2024. Why has advanced commercial HVAC control not yet achieved its promise? \textit{Journal of Building Performance Simulation} 18(2):217--228.
\bibitem[Hersbach et al.(2020)]{hersbach2020} Hersbach, H., B. Bell, P. Berrisford, et al. 2020. The ERA5 global reanalysis. \textit{Quarterly Journal of the Royal Meteorological Society} 146(730):1999--2049.
\bibitem[Khabbazi and Kircher(2026)]{khabbazi2026acc} Khabbazi, A.J., and K.J. Kircher. 2026. Small HVAC control demonstrations in larger buildings often overestimate savings. \textit{Proceedings of the American Control Conference}. arXiv:2602.22499.
\bibitem[Khabbazi et al.(2024)]{khabbazi2024} \new{Khabbazi, A.J., E.N. Pergantis, L.D. Reyes Premer, A.H. Lee, J. Ma, H. Liu, G.P. Henze, and K.J. Kircher. 2024. What have we learned from field demonstrations of advanced commercial HVAC control? \textit{Proceedings of the 8th International High Performance Buildings Conference at Purdue}, West Lafayette, IN, Paper 473.}
\bibitem[Khabbazi et al.(2025)]{khabbazi2025} Khabbazi, A.J., E.N. Pergantis, L.D. Reyes Premer, P. Papageorgiou, A.H. Lee, J.E. Braun, G.P. Henze, and K.J. Kircher. 2025. Lessons learned from field demonstrations of model predictive control and reinforcement learning for residential and commercial HVAC: A review. \textit{Applied Energy} 370:126459.
\bibitem[Kircher and Zhang(2015)]{kircher2015} \new{Kircher, K.J., and K.M. Zhang. 2015. On the lumped capacitance approximation accuracy in RC network building models. \textit{Energy and Buildings} 108:454--462.}
\bibitem[Liu and Henze(2006)]{liu2006} Liu, S., and G.P. Henze. 2006. Experimental analysis of simulated reinforcement learning control for active and passive building thermal storage inventory: Part 2: Results and analysis. \textit{Energy and Buildings} 38(2):148--161.
\bibitem[Pergantis et al.(2024a)]{pergantis2024} Pergantis, E.N., Priyadarshan, N. Al Theeb, P. Dhillon, J.P. Ore, D. Ziviani, E.A. Groll, and K.J. Kircher. 2024. Field demonstration of predictive heating control for an all-electric house in a cold climate. \textit{Applied Energy} 360:122820.
\bibitem[Pergantis et al.(2024b)]{pergantis2024humidity} \new{Pergantis, E.N., P. Dhillon, L.D. Reyes Premer, A.H. Lee, D. Ziviani, and K.J. Kircher. 2024. Humidity-aware model predictive control for residential air conditioning: A field study. \textit{Building and Environment} 266:112093.}
\bibitem[Ruud et al.(1990)]{ruud1990} Ruud, M.D., J.W. Mitchell, and S.A. Klein. 1990. Use of building thermal mass to offset cooling loads. \textit{ASHRAE Transactions} 96(2).
\bibitem[Serale et al.(2018)]{serale2018} Serale, G., M. Fiorentini, A. Capozzoli, D. Bernardini, and A. Bemporad. 2018. Model predictive control (MPC) for enhancing building and HVAC system energy efficiency: Problem formulation, applications and opportunities. \textit{Energies} 11(3):631.
\bibitem[Zanetti et al.(2025)]{zanetti2025} \new{Zanetti, E., D. Blum, H. Fu, C. Weyandt, M. Pritoni, and M.A. Piette. 2025. Commercial building HVAC demand flexibility with model predictive control: Field demonstration and literature insights. \textit{Energy and Buildings} 345:116097.}
\end{thebibliography}
\end{document}